\documentclass[fleqn,10pt]{wlscirep}
\usepackage[utf8]{inputenc}
\usepackage[T1]{fontenc}
\usepackage{subfigure}
\usepackage{float}

\title{Parametric FEM simulation of composite barrier FTJs under external 
	bias at room temperature}

\author[1]{Catalin Tibeica}
\author[1]{Titus Sandu}
\author[1]{Oana Nedelcu}
\author[1]{Rodica Plugaru}
\author[1,*]{Neculai Plugaru}
\affil[1]{National Institute for Research and Development in Microtechnologies (IMT - 
	Bucharest), 126A, Erou Iancu Nicolae Street, Voluntari Ilfov, 077190, 
	Romania}

\affil[*]{neculai.plugaru@imt.ro}

\begin{abstract}
A study on a parametrized model of a composite barrier FTJ (three-interface system, with a non-polar dielectric layer) 
under an external bias voltage and at room temperature, using FEM-based simulations, was performed. The approach 
involves the Thomas-Fermi model assuming incomplete screening of polarization charges for building the energy barrier 
profile, and numerically simulates the electron transport through the barrier by bias-voltage-dependent tunneling, using 
Tsu-Esaki formulation. That naturally include the temperature dependent contributions to the total current density. The 
TER coefficient and current densities are computed considering variation of a large set of parameters that describe the 
composite barrier FTJ system in realistic physical range of values with respect to a reference (prototypical) system. In 
this study, the parametric simulations were performed starting from selected data reported on the SRO/STO/BTO/SRO 
heterostructure. The most important results of our work can be stated as follows: \textit{i)} The FEM simulations 
prove to be reliable approach when we are interested in the prediction of FTJ characteristics at temperatures close to 
300 K, and \textit{ii)} We show that several configurations with large TER values may be predicted, but at the 
expense of very low current densities in the ON state. We suggest that the results may be useful for assessing the FTJ 
performances at ambient temperature, as well as to design preoptimized FTJs by using different combinations of materials 
to comply with a set of properties of a specific model.
\end{abstract}
\begin{document}

\flushbottom
\maketitle
%
%
\thispagestyle{empty}


\section*{Introduction}
The ferroelectric tunneling junction (FTJ) device proposed by Esaki 
\textit{et al.} as a \textit{polar switch } in 1971 \cite{Esaki1971}, has marked the beginning of thorough theoretical 
and experimental researches on understanding the physical mechanisms involved, as well as 
improving the predictive models to assist further developments of new FTJs 
applications. Basically, a FTJ design consists of two electrodes separated 
by a nanometer-thick ferroelectric layer, and its operation relies on 
switching between two resistive states when the ferroelectric polarization 
changes its direction normal to the junction plane. More complex designs 
include additional layers for improving the overall performance \cite{Zhuravlev2009, Garcia2014, Velev2016, He2019, Yang2019, Hwang2021}.

In terms of modelling a FTJ structure, the \textit{ab-initio} methods provide the most 
detailed insight into the local structure and physics at atomic level, but 
usually they are computationally consuming. On the other hand, the continuum 
medium model approaches, in which the numerical solution of Schr\"{o}dinger 
equation is solved by Finite Element Method (FEM) can provide meaningful 
information on the FTJs behavior, at low computational costs, despite their 
idealization at nanometer scale, where the discrete behavior of nature 
manifests.

Nowadays, there is a large variety of materials to build FTJ 
heterostructures therefore predicting the performance of various designs or 
finding optimal structures at reduced costs becomes of real interest prior 
to experiments. Nevertheless, realistic continuum medium models are 
described by a relatively large number of parameters, which include 
thickness dependent material properties, and material to material interface 
properties, many of them being estimated from data obtained by \textit{ab-initio} 
calculations. Finding complete sets of data for many materials of interest 
for the FTJ simulation in a FEM framework has proven a challenging task.

Mostly in theoretical studies, the FTJ performance is asserted by the 
tunneling electroresistance ratio (TER) expressed as (G$_{high}$ - 
G$_{low})$ / G$_{low}$, where G$_{high / low }$ are the conductivities in the 
low/high resistance states at 0 K temperature. TER is an overall merit 
parameter and only its value cannot state the practical usefulness of a FTJ 
for applications because the magnitude of the write and readout currents in 
the ON and OFF states is also of utmost importance. 

Starting from these considerations, in this work we have carried out 
FEM-based simulations on a parametrized FTJ model with a composite barrier 
including a non-polar dielectric layer besides the ferroelectric one, as an 
additional handle for the FTJ performance enhancement. Usually, in the case 
of a three-interface system, large TER values may be obtained, but they 
result from very low current density in the ON state. Predicting the current 
densities through the junction requires bias-voltage-dependent tunneling 
simulations which allow to evaluate the TER values as $(j_{ON} - j_{OFF}) / j_{OFF}$. To achieve this aim, we use a 
well-established continuum medium model for such structures, which assumes the incomplete 
screening of the polarization charges within the Thomas-Fermi approximation 
to build the energy barrier profile \cite{Zhuravlev2009, Mehta1973}, and simulate the electron 
transport through the barrier by bias-voltage-dependent tunneling at room 
temperature \cite{He2019, Franchini2020}.

The paper is organized as follows. In Section 2 we define the parameters of 
the model (Subsection 2.1) and briefly review the theoretical background of 
our approach (Subsection 2.2). The simulation results of the reference model 
used in this work, including potential barriers, transmission coefficients, 
current densities and TER values versus $V_{a}$ are presented in Subsection 
2.3. The results of the parametric simulations are shown in Section 3, and 
their discussion is presented in Section 4. Section 5 highlights the main 
findings and outreach of this work.

\section*{Model and simulation}

The numerical simulations were performed by using the COMSOL Multiphysics 
software \cite{comsol}. Two main steps must be followed when simulating a 
FTJ in the frame of a continuum medium model, namely the construction of the 
energy barrier, and the simulation of electron transport process through the 
barrier. These determine the current densities corresponding to the two 
polarization directions in the ferroelectric, and hence the TER coefficient 
is derived.

\subsection*{Building the energy barrier}

\begin{figure}[H]
\centering
  \includegraphics[width=4in,height=3in]{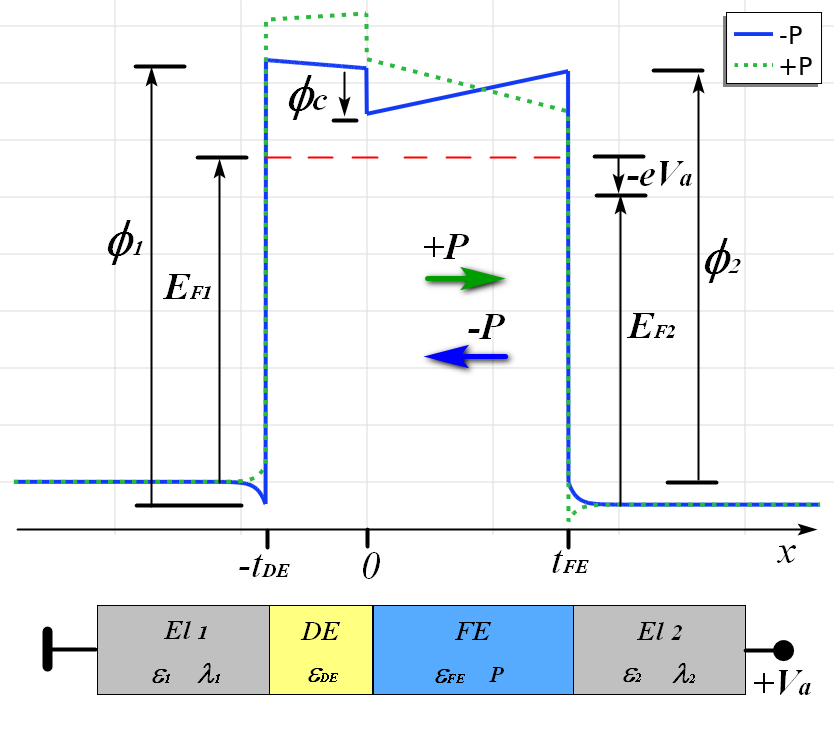}
\caption{ Structure and energy band diagram of a FTJ system with 3 
interfaces: $\lambda _{1,2}$ are the Thomas-Fermi screening lengths in the 
electrodes, ${\varepsilon}_{1}$, ${\varepsilon}_{2}$, ${\varepsilon}_{DE}$, ${\varepsilon}_{FE}$ stand for the relative 
dielectric constants of the electrodes, dielectric and ferroelectric, 
respectively, $t_{DE}$ and $t_{FE}$ stand for the thicknesses of the dielectric 
and ferroelectric layers, $\Phi _{1}$ and $\Phi _{2}$ denote the conduction band discontinuities at El1/DE and El2/FE 
interfaces, \textit{$\Phi $}$_{c}$ is the band offset at the DE/FE interface, $E_{F1}$ and $E_{F2}$ are the Fermi 
energies in electrodes, $P$ is the ferroelectric polarization, and $V_{a}$ is the external applied potential.}
\label{fig1}
\end{figure}

The key property responsible for the FTJ functionality is the change in 
energy barrier profile upon the ferroelectric polarization direction, as 
illustrated in Figure \ref{fig1}. The electron transport through the asymmetric barrier 
defines its electrical conductance with respect to the polarization 
direction, and therefore the junction can be switched between two states (ON 
for the higher conductance, and OFF for the lower conductance).

The electrostatic potential across a FTJ is determined by three components, 
namely the built-in electric field, $V_{Bi}$, due to different workfunctions 
and chemistry at interfaces, the external applied electric field, $V_{a}$, 
and the depolarization field, $V_{d}$, caused by the incomplete screening of 
polarization charges at electrodes. The conduction band offsets at 
interfaces have also to be considered in order to build the energy barrier, 
but generally their values are not well known \cite{Junquera2008, Stengel2011}.

The potential energy of the barrier as seen by an electron, $U(x)$, is defined on 
4 intervals, corresponding to the spatial domain occupied by each material 
in the system. We followed the derivation of the potential energy as can be 
found in various papers \cite{He2019, Chang2017, Su2022}, but only its expression is given here, 
using our notations:

\begin{equation}
	\label{eq1}
	U\left( x \right) = \left\{ {\begin{array}{l}
			 e\frac{\sigma _s \lambda _1 }{\varepsilon _1 \varepsilon _0 }\exp \left[ {(x+t_{DE}) 
				\mathord{\left/ {\vphantom {(x+t_{DE}) {\lambda _1 }}} \right. 
					\kern-\nulldelimiterspace} {\lambda _1 }} \right],\;x < - t_{DE} \\ 
			 e\sigma _s \left( {\frac{\lambda _1 }{\varepsilon _1 \varepsilon _0 } + 
				\frac{x+t_{DE} }{\varepsilon _{DE} \varepsilon _0 } } \right) + \varPhi _1 ,\; - t_{DE} \le x < 0 \\ 
			 e\sigma _s \left( {\frac{\lambda _1 }{\varepsilon _1 \varepsilon _0 } + 
				\frac{t_{DE} }{\varepsilon _{DE} \varepsilon _0 }} \right) + e\frac{\sigma _s 
				- P}{\varepsilon _{FE} \varepsilon _0 }x + \varPhi _1 - \varPhi _C ,\;0 \le 
			x \le t_{FE} \\ 
			-e\left( {V_a + V_{BI} + \frac{\sigma_s \lambda _2 }{\varepsilon _2 \varepsilon _0 
				}\exp \left[ { - {\left( {x - t_{FE} } \right)} \mathord{\left/ {\vphantom 
							{{\left( {x - t_{FE} } \right)} {\lambda _2 }}} \right. 
						\kern-\nulldelimiterspace} {\lambda _2 }} \right]} \right) + \varPhi _1 - \varPhi _C - \varPhi _2 ,{\kern 1pt} {\kern 1pt} x > t_{FE} 
			\\ 
	\end{array}} \right.
\end{equation}

\noindent
where $\sigma _{s}$ is the screening charge per unit area at the electrode 
interfaces, $\lambda _{1,2}$ are the Thomas-Fermi screening lengths in 
electrodes, P is the ferroelectric polarization, $\Phi _{i}$ are the band 
discontinuities at the interfaces, V$_{a}$ is the external applied 
potential, $V_{Bi} = ({\Phi }_{2} + {\Phi }{c} - {\Phi }_{1} - E_{F2}+ E_{F1})$ is the built-in 
potential, and $t_{FE/DE}$ are the layers thicknesses.

\subsection*{Tunneling through the energy barrier}

Maybe the most used method for calculating the transmission coefficient 
across energy barriers of FTJs is the WKB (Wentzel-Kramers-Brillouin) 
approximation, which is a semiclassical approach to compute the stationary 
solution of the Schr\"{o}dinger equation, with the barrier approximated by 
its average value \cite{Velev2016, Ma2019, Gruverman2009}. However, the WKB approximation cannot explain 
the resonance phenomena appearing in some systems \cite{Sandu2022} and is inaccurate for 
barrier profiles which vary abruptly. Therefore, alternative methods were 
proposed for calculating the transmission current across arbitrary potential 
barriers \cite{Ando1987}.

In the Tsu-Esaki model \cite{Tsu1973, Tuomisto2017}, the net current through the barrier is 
derived considering only the longitudinal wavevector of electrons in the 
tunneling process, and it is also assumed that the dispersion relation, 
$k(E)$, is parabolic in the electrodes

\begin{equation}
\label{eq2}
j= \frac {4\pi\; e \: m^* k_B T}{{h^3}} \int_0^\infty {D(E) \ln \left\{ \frac {1+\exp \left [ 
      \frac{E_{F1}-E}{k_B T}\right]}{1+\exp \left [\frac{E_{F2} - E - eV_a}{k_B T} \right ]} \right\} dE}
\end{equation}

\noindent
where $m^{\ast }$ is the electron effective mass, $E_{F1 / F2}$ are the Fermi 
energies in the electrodes, $V_{a}$ is the bias voltage across the 
heterojunction, $D(E)$ is the transmission coefficient through the barrier as 
function of energy, and $E$ is the wavevector energy component perpendicular to 
the interface.

The term $k_B T\mbox{ln}\left\{ {\frac{\mbox{1} + \mbox{exp}\left[ 
{\frac{E_F - E}{k_B T}} \right]}{\mbox{1} + \mbox{exp}\left[ {\frac{E_F - E 
- eV_a }{k_B T}} \right]}} \right\}$ is known as the supply function \cite{Tsu1973, Duan2020} and is derived from energy distributions of electrons in electrodes, in 
Fermi-Dirac statistics. As one can see, the above relations include both the 
temperature, $T$, and externally applied bias voltage, $V_{a}$, as parameters.

The Tsu-Esaki approach, by means of relation (\ref{eq2}), is used in our work. The 
upper limit of the integral in relation (\ref{eq2}) was set higher than the maximum 
barrier height, so the computed current density virtually includes the 
Fowler-Nordheim (tunneling across a triangular barrier) and thermionic 
tunnelling mechanisms \cite{Guo2020}.

\subsection*{Simulation of the reference system}
Because finding complete sets of data on the materials involved in a FTJ 
system is a challenging task, we have parameterized the model as explained 
in Section 3. Following similar approaches, where the analysis is built 
around a reference or prototypical FTJ system \cite{He2019, Su2022, Banerjee2019, Klyukin2018}, herein we 
consider the SRO/STO/BTO/SRO heterostructure as a reference, with the 
characteristic parameters listed in Table~\ref{tab1}.

The energy band diagram of the analyzed system is analytically implemented 
in the program based on relations (\ref{eq1}) and then used in a numerical model 
that solves the single-particle (electron) one-dimensional Schr\"{o}dinger 
equation. The transmission coefficient through the barrier as a function of 
energy, $D(E)$, is computed using a stationary study for both directions of the 
ferroelectric polarization. Then, for each analyzed system, the tunneling 
current densities are computed using the relation (\ref{eq2}) and the TER 
coefficient is derived as TER = ($j_{ON}$ -- $j_{OFF})$ / $j_{OFF}$.

\begin{table}[H]
\begin{tabular}
{|p{44pt}|p{47pt}|p{47pt}|p{47pt}|p{47pt}|p{47pt}|p{47pt}|}
\hline
& 
$\varepsilon _{r}$& 
m* [m$_{0}$]& 
E$_{F}$ [eV]& 
$\Phi $[eV]& 
$\lambda $[nm]& 
P [C/m$^{2}$] \\
\hline
\textit{SrRuO}$_{3}$& 
8.45& 
5& 
3& 
\raisebox{-1.50ex}[0cm][0cm]{3.6}& 
0.075& 
- \\
\cline{1-4} \cline{6-7} 
\raisebox{-1.50ex}[0cm][0cm]{\textit{SrTiO}$_{3}$}& 
\raisebox{-1.50ex}[0cm][0cm]{300}& 
\raisebox{-1.50ex}[0cm][0cm]{2}& 
\raisebox{-1.50ex}[0cm][0cm]{-}& 
 & 
\raisebox{-1.50ex}[0cm][0cm]{-}& 
\raisebox{-1.50ex}[0cm][0cm]{-} \\
\cline{5-5} 
 & 
 & 
 & 
 & 
\raisebox{-1.50ex}[0cm][0cm]{0}& 
 & 
  \\
\cline{1-4} \cline{6-7} 
\raisebox{-1.50ex}[0cm][0cm]{\textit{BaTiO}$_{3}$}& 
\raisebox{-1.50ex}[0cm][0cm]{125}& 
\raisebox{-1.50ex}[0cm][0cm]{2}& 
\raisebox{-1.50ex}[0cm][0cm]{-}& 
 & 
\raisebox{-1.50ex}[0cm][0cm]{-}& 
\raisebox{-1.50ex}[0cm][0cm]{0.16} \\
\cline{5-5} 
 & 
 & 
 & 
 & 
\raisebox{-1.50ex}[0cm][0cm]{3.6}& 
 & 
  \\
\cline{1-4} \cline{6-7} 
\textit{SrRuO}$_{3}$& 
8.45& 
5& 
3& 
 & 
0.075& 
- \\
\hline
\end{tabular}
\caption{Parameters of the reference system SRO/STO/BTO/SRO: 
	relative dielectric constant, $\varepsilon _{r}$, electron effective mass, 
	$m*$, Fermi energy, E$_{F}$, band discontinuity, $\Phi $, Thomas-Fermi 
	screening length, $\lambda $, ferroelectric polarization, $P$.}
\label{tab1}
\end{table}

A particular feature of the SRO/STO/BTO/SRO system is that the band offset 
at the dielectric/ferroelectric interface is null, and the built-in voltage 
is also null because the Fermi energies are equal ($E_{F1}=E_{F2})$ and 
the electrodes band discontinuities are equal (\textit{$\Phi $}$_{1}$ = \textit{$\Phi $}$_{2})$.

Using the data in Table \ref{tab1} for the reference system, and assuming constant 
values for the dimensional and external parameters ($t_{EL1}=t_{EL2}$ = 4 
nm, $t_{DE}$ = 2 nm, $t_{FE}$ = 4 nm, $V_{a}$ = 0.005 V, $T$ = 300 K), the 
calculated TER$_{ref}$ = 37.4. To ensure the accuracy of the results in 
solving the Schr\"{o}dinger equation, a convergence analysis of the 
discretization in space and energy domains was performed, step sizes of 
0.0025 nm, and 0.001 eV respectively, were determined as optimal and 
henceforth used in all simulations.

\begin{figure}[H]
\centering
\includegraphics[width=4in,height=3in]{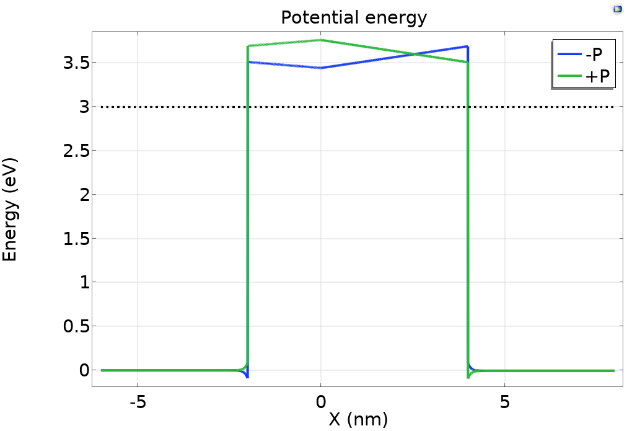}
\caption{Barrier shape of the reference FTJ system, seen by an 
	electron, with respect to the ferroelectric polarization directions. For +$P $ the 
	average height of the barrier is bigger than for -$P$ .}
\label{fig2}
\end{figure}

\begin{figure} [H]
\centering
\includegraphics[width=4in,height=3.00in]{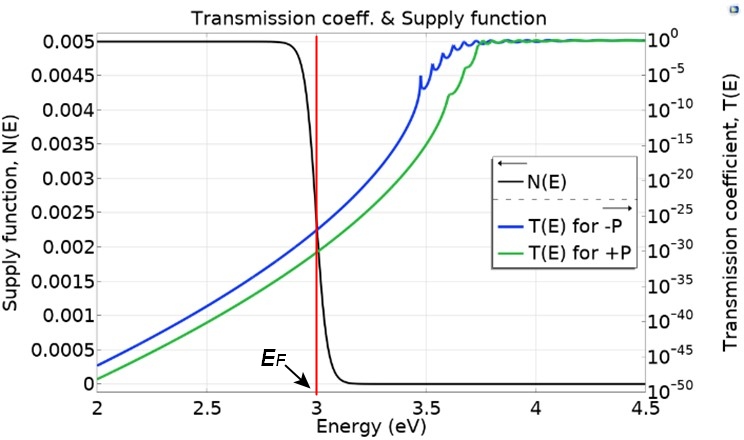}
\caption{Supply function, $N(E),$ as function of electron energy, for 
	$E_{F}$ = 3 eV, V$_{a}$ = 0.005 V and $T$ = 300 K (left-side scale), and 
	transmission coefficient through the barrier, $T(E)$, for -$P$ and +$P$ polarization 
	of the ferroelectric, as function of electron energy (right-side scale, 
	logarithmic).}
\label{fig3}
\end{figure}

The relevant characteristics of the reference system under simulation are 
shown in Figures \ref{fig2} and  \ref{fig3}. They include the barrier profiles for both 
polarization directions, the supply function, $N(E)$, and the transmission 
coefficient in tunneling of the electrons through the barrier.

\begin{figure}[H]
\centering
\includegraphics[width=4.00in,height=3.00in]{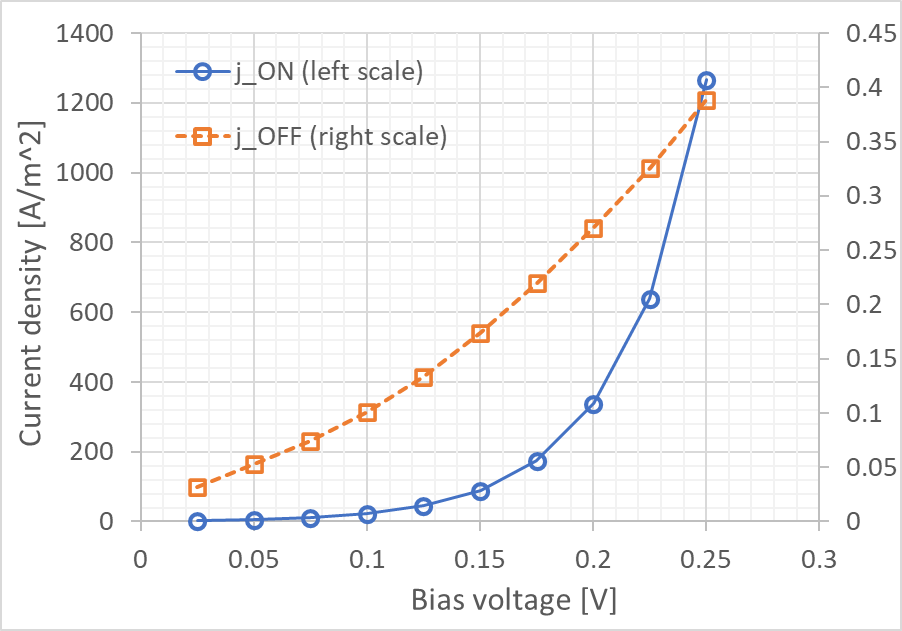}
\caption{Current densities versus $V_{a}$ for the reference FTJ 
	system.}
\label{fig4}
\end{figure}

\begin{figure}[H]
\centering
\includegraphics[width=4.00in,height=3.00in]{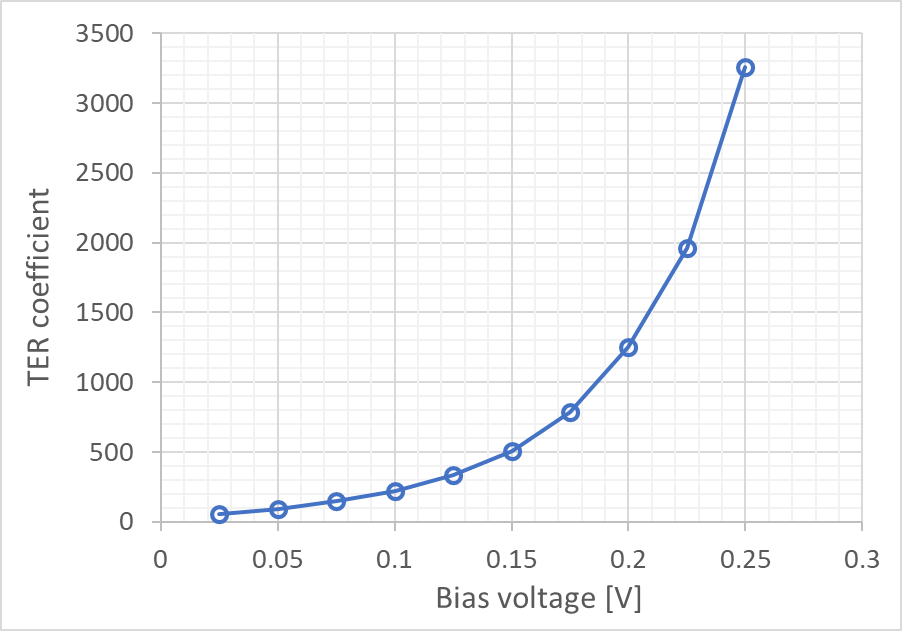}
\caption{TER coefficient versus $V_{a}$ for the reference FTJ system.}
\label{fig5}
\end{figure}

By considering the external bias voltage as a variable parameter in 
simulations, $V_{a}$ = 0.025--0.3 V, $j(V_{a})$ and TER($V_{a})$ curves are also 
obtained (Figure \ref{fig4} and \ref{fig5}). However, in the following parametric analyses, we 
set a fixed lower bias voltage (0.005 V) to keep the Fowler-Nordheim 
contribution at a low value and thus to highlight the thermal effects that 
characterize the FTJs behavior at ambient temperature.

\section*{Parametric simulations}

Once the reference system was fully characterized, we varied the parameters 
that describe it within limits which encompass realistic ranges of values of 
a FTJ design. We also note that the variation of a given parameter was done 
around the reference value and independently of the other parameters.

The full description of a three-interface FTJ model requires a series of 
parameters used for barrier construction and transport (tunneling) problems. 
First, one must consider the material and interface properties, as follows:

\begin{itemize}
\item the electric polarization, $P$, of the ferroelectric film, considered 
perpendicular to the FTJ plane in all simulations in this work,
\item the Thomas-Fermi screening lengths, \textit{$\lambda $}$_{1}$ and \textit{$\lambda $}$_{2}$, of the electrodes,
\item the dielectric constants, \textit{$\varepsilon $}$_{r}^{El1}$, \textit{$\varepsilon $}$_{r}^{El2}$, $\varepsilon 
_{r}^{FE}$, $\varepsilon _{r}^{DE}$, of the electrodes, 
ferroelectric, and dielectric materials, respectively,
\item Fermi energy of the electrodes, $E_{F1}$ and $E_{F2}$ ,
\item band offsets at interfaces, \textit{$\Phi_1$}, \textit{$\Phi_2$} and \textit{$\Phi_c$}, or the 
barriers instead, as $U_d = \Phi_1 - E_{F1}$, \ $U_f = \Phi_2 - E_{F2}$ ,
\item electron effective masses, $m*_{El1}$, $m*_{El2}$, $m*_{FE}$, $m*_{DE}$, in the 
electrodes, ferroelectric and dielectric materials, respectively.
\end{itemize}

Therefore, to characterize the performance of a certain FTJ in the frame of 
this model, one needs to know the values of 16 material and interface 
parameters (listed in Table 1). In the case that the electrodes consist of 
the same material, the number of parameters reduces to 12, since 
\textit{$\varepsilon $}$_{r}^{El1}$\textit{ = $\varepsilon $}$_{r}^{El2}$, $E_{F1} = E_{F2}$, \textit{$\lambda 
$}$_{1}$\textit{ = $\lambda $}$_{2}$, and $m*_{El1}$ = $m*_{El}$. Apparently, the Fermi energy should be a parameter 
to be considered variable in our model, but it turns out just as a reference level. Moreover, 
in this work we deal only with the parameters that describe the barrier 
shape, so the effective masses of electrons are not varied, which means that 
only 7 parameters remain. The dimensional parameters, and $V_{a}$ are 
separately treated for a given system. The temperature of the simulated 
system is always considered constant, T = 300 K, if not otherwise specified.

In the following, we present the results of the simulations performed by 
varying the above-mentioned seven material-based parameters. Lastly, we 
include the thickness of the dielectric layer, t$_{DE}$, in our analysis, as 
a dimensional parameter which plays the most important part in the composite 
barrier FTJs \cite{Garcia2014, Velev2016, He2019, Su2022}. On the graphs representing the TER coefficient 
versus variable parameter, the position of the reference system is 
graphically marked by a ``\textbf{+}'' sign and represents the previously 
calculated value, TER$_{ref}$ = 37.4.

\subsection*{Effect of ferroelectric polarization, $P$}
\begin{figure}[htbp]
	\begin{center}
		\subfigure {\label{fig6a} \includegraphics [width=2.5in,height=2.5in]{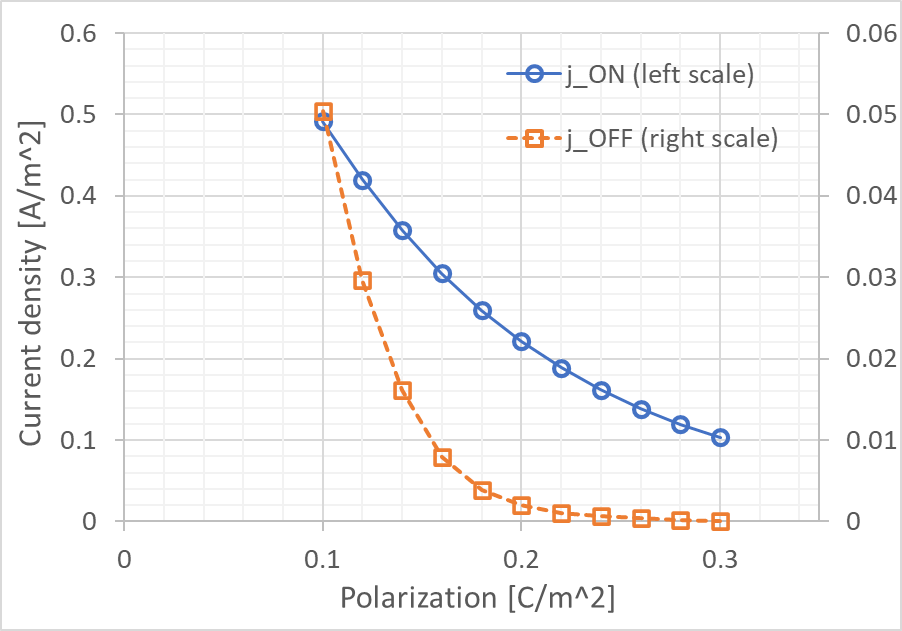}}  
		\subfigure {\label{fig6b} \includegraphics [width=2.50in,height=2.5in]{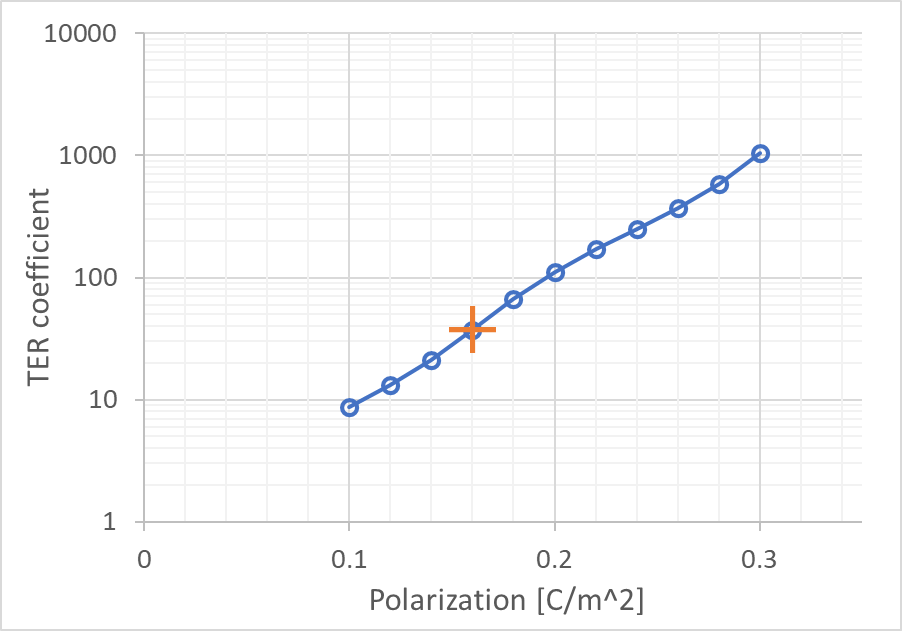}} 
	\end{center}
	\caption{(a)  Current densities versus $\vert $P$\vert $ (b)TER coefficient versus $\vert $P$\vert $.}
	\label{fig6}
\end{figure}

The current densities versus polarization curves are plotted in Figure \ref{fig6}(a), 
and the resulting TER values are shown in Figure \ref{fig6}(b). One may note that a 
small variation in $P$, from 0.1 to 0.3 C/m$^{2}$ determines a large TER 
variation, of two orders of magnitude. However, the j$_{ON}$ value for $P$ = 
0.3 C/m$^{2}$, at maximum TER, is only 0.1 A/m$^{2}$ (10$^{ - 19}$ 
A/nm$^{2})$, which could result in small read currents for the sensing 
circuitry when scaling down the device.

\subsection*{Effect of Thomas-Fermi length and electrode permittivity, $\lambda$/$\varepsilon_{1}$}

\begin{figure}[htp]
	\begin{center}
		\subfigure {\label{fig7a} \includegraphics [width=2.5in,height=2.5in]{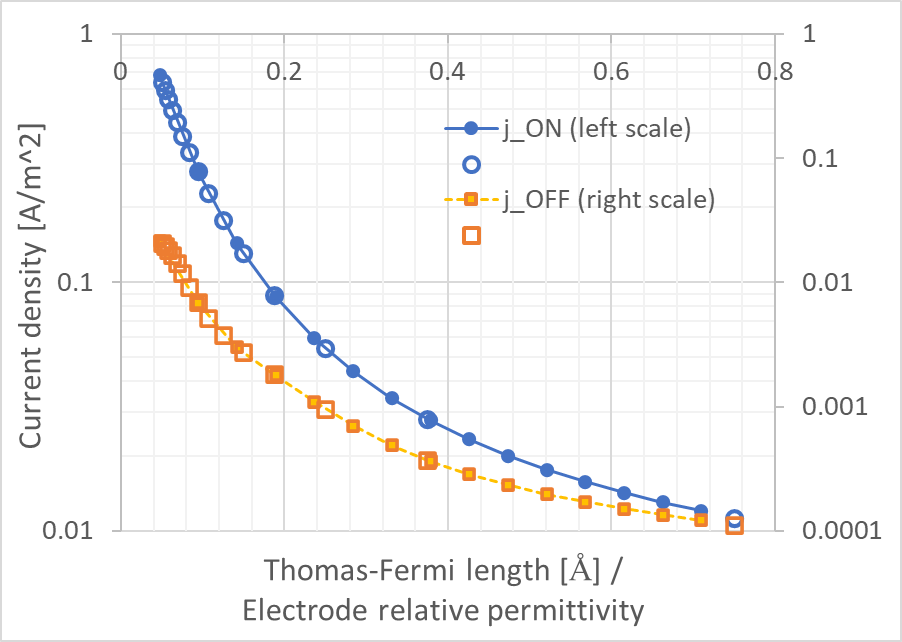}}  
		\subfigure {\label{fig7b} \includegraphics [width=2.50in,height=2.5in]{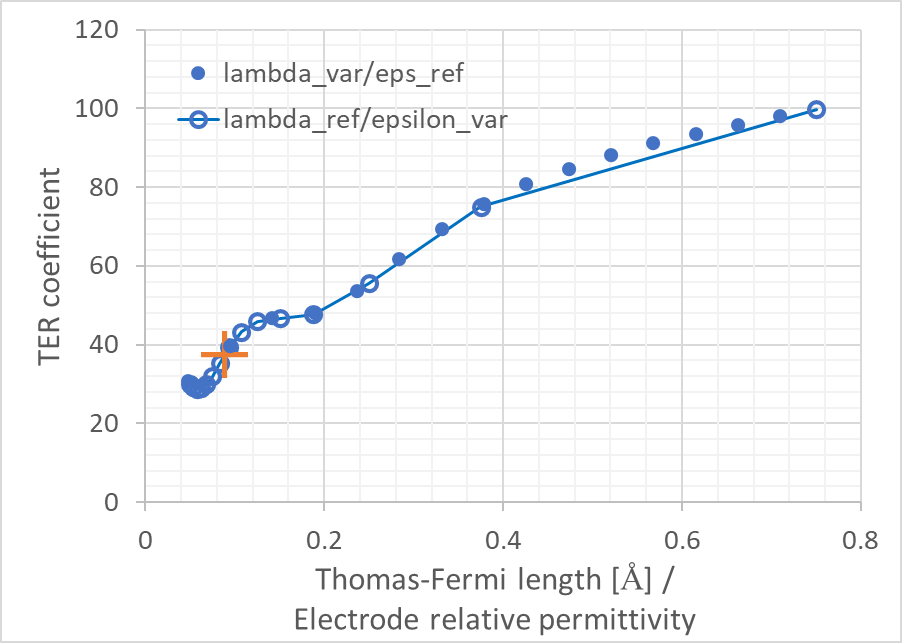}} 
	\end{center}
	\caption{(a) Current densities versus \textit{$\lambda $/}\textit{$\varepsilon $}$_{1}$ (b) TER coefficient versus 
\textit{$\lambda $/}\textit{$\varepsilon $}$_{1}$.}
	\label{fig7}
\end{figure}

One may note in Eq. (\ref{eq1}) that $U(x)$ in the barrier region is dependent on 
\textit{$\lambda $/$\varepsilon $}$_{1}$, and hence, the representation as a function of this ratio is more 
meaningful than using individual parameters. Thus, taking \textit{$\lambda $/$\varepsilon $}$_{1}$ as a 
parameter, we obtain the $j_{ON / OFF}$ and TER plots shown in Figure \ref{fig7}(a) and 
(b). From these data it may be derived that one has to do a trade-off 
between \textit{$\lambda $} and \textit{$\varepsilon $}$_{1}$ values in order to obtain $j_{ON}$ and TER values that 
are adequate for applications.

\subsection*{Effect of dielectric permittivity, $\varepsilon_{DE}$}

\begin{figure}[htp]
	\begin{center}
		\subfigure {\label{fig8a} \includegraphics [width=2.5in,height=2.5in]{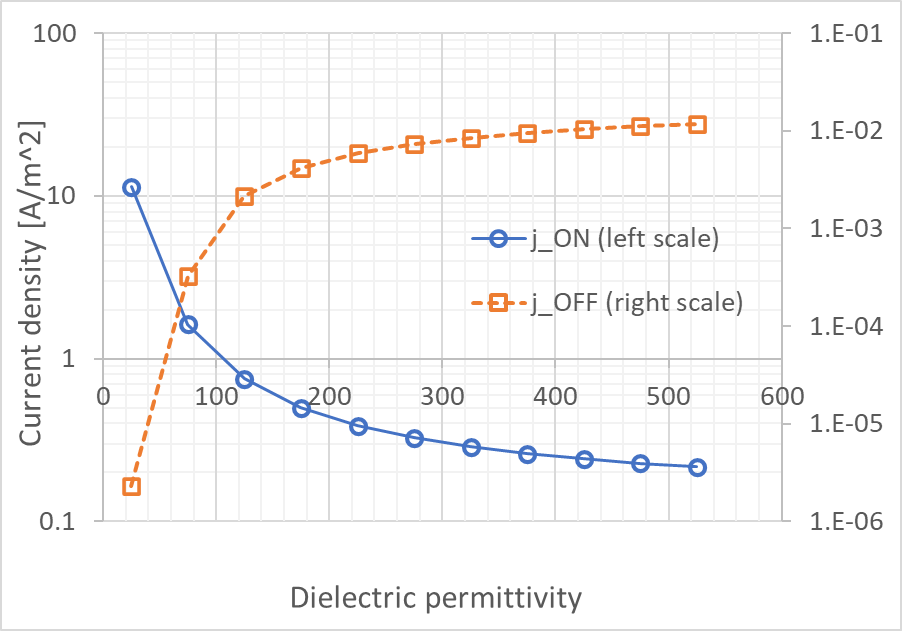}}  
		\subfigure {\label{fig8b} \includegraphics [width=2.50in,height=2.5in]{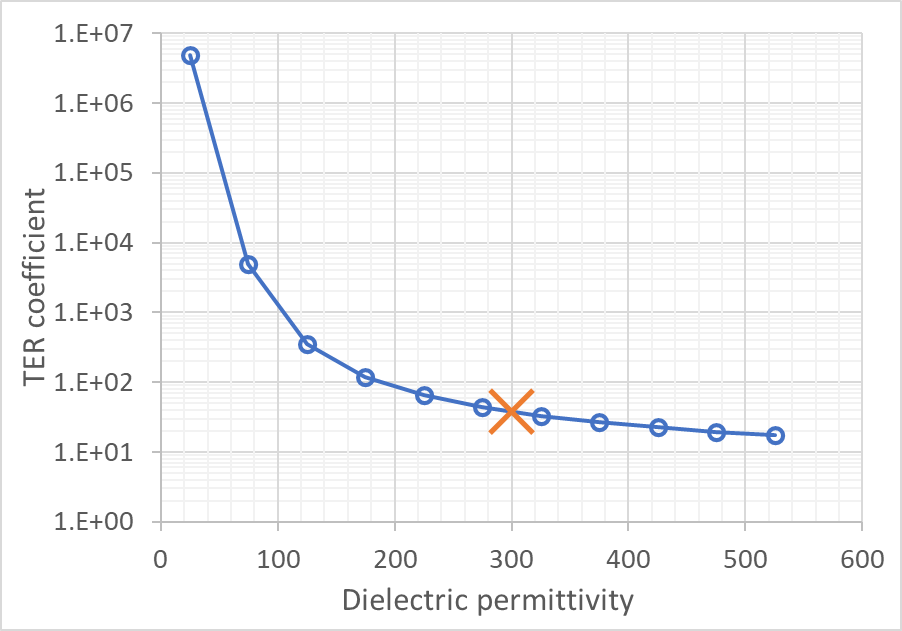}} 
	\end{center}
	\caption{(a) Current densities versus \textit{$\varepsilon $}$_{DE}$ (b) TER coefficient versus \textit{$\varepsilon 
$}$_{DE}$.}
	\label{fig8}
\end{figure}

Both the current densities and the TER ratio show a strong dependence on 
\textit{$\varepsilon $}$_{DE}$, for a given $t_{DE}$ value (see Figures \ref{fig8}(a) and (b)). The plots 
suggest that both $j_{ON}$ and TER may acquire large values when 
\textit{$\varepsilon $}$_{DE}$ is less than 100, in the present case.

\subsection*{Effect of ferroelectric permittivity, $\varepsilon_{FE}$}

\begin{figure}[htp]
	\begin{center}
		\subfigure {\label{fig9a} \includegraphics [width=2.5in,height=2.5in]{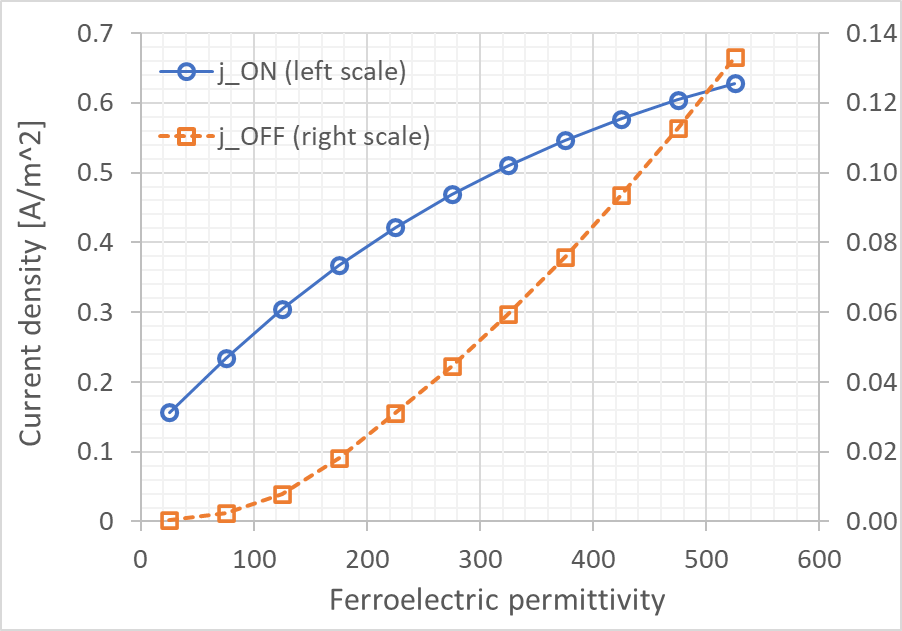}}  
		\subfigure {\label{fig9b} \includegraphics [width=2.50in,height=2.5in]{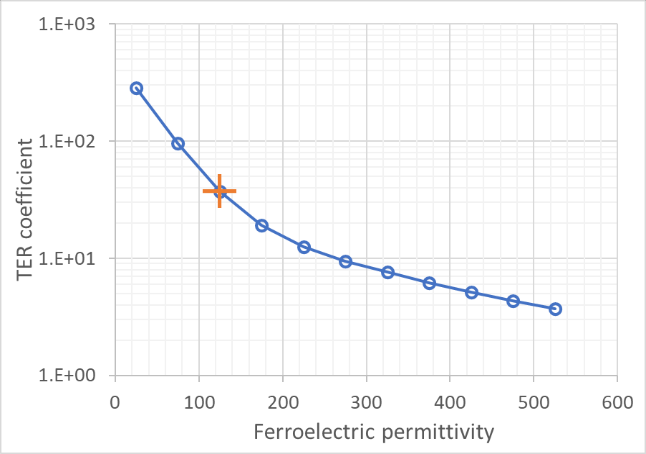}} 
	\end{center}
	\caption{(a) Current densities versus \textit{$\varepsilon $}$_{FE}$ (b) TER coefficient versus \textit{$\varepsilon 
$}$_{FE}$.}
	\label{fig9}
\end{figure}

Similar with the $\varepsilon _{DE}$ effect discussed at 3.3, the plots in 
Figures \ref{fig9}(a) and (b) show that $\varepsilon _{FE}$ parameter has a 
significant impact on the FTJ performance. The j$_{ON}$ and TER versus 
$\varepsilon _{FE}$ curves suggest an optimum range of the ferroelectric 
permittivity between about 50 and 200.

\subsection*{Effect of electrode-dielectric (ferroelectric) barrier, $ U_{d(f)}$}

\begin{figure}[H]
	\begin{center}
		\subfigure {\label{fig10a} \includegraphics [width=2.5in,height=2.5in]{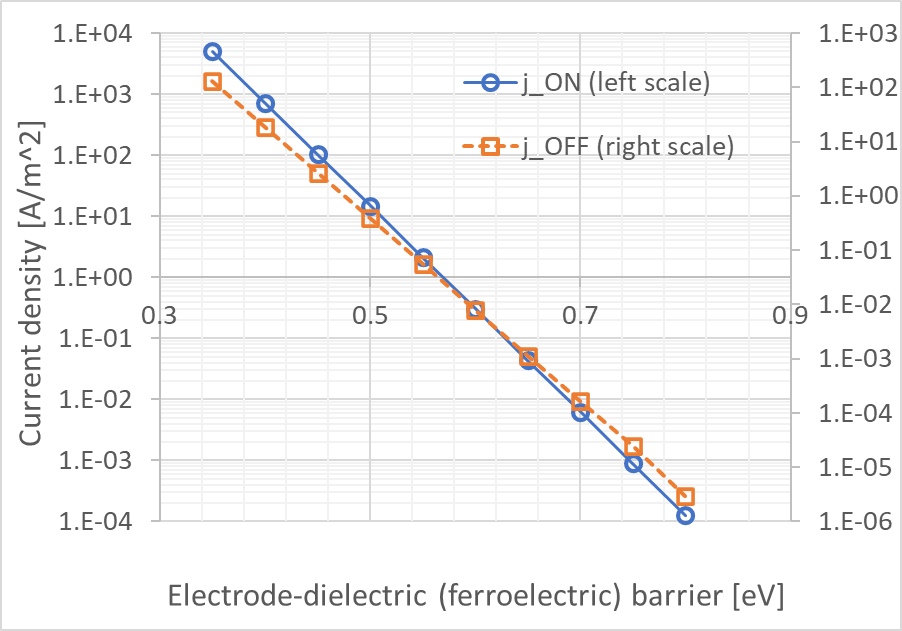}}  
		\subfigure {\label{fig10b} \includegraphics [width=2.50in,height=2.5in]{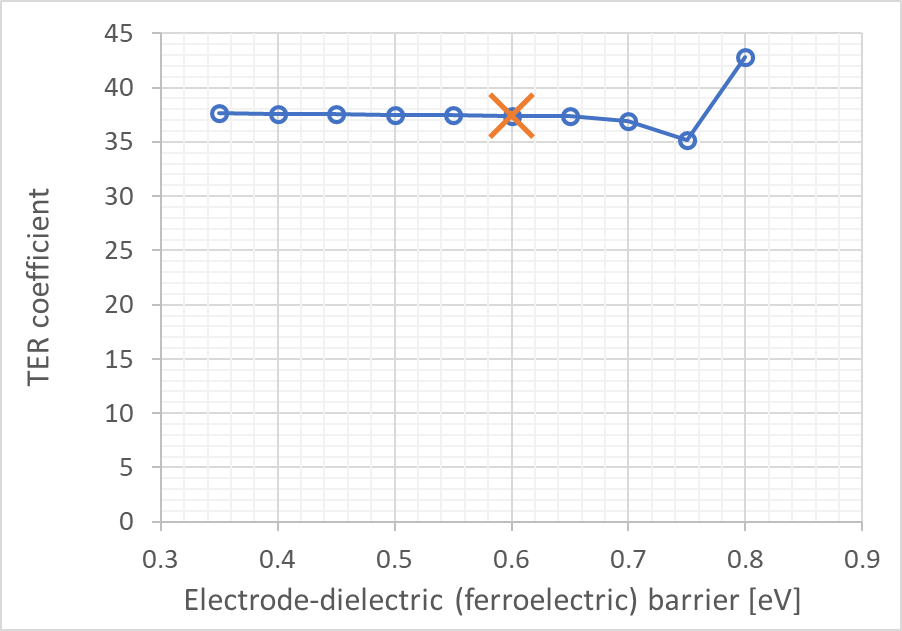}} 
	\end{center}
	\caption{(a) Current densities versus $U_{d(f)}$ (b) TER coefficient versus $U_{d(f)}$.}
	\label{fig10}
\end{figure}

Although each of the current densities vary about eight orders of magnitude 
when the energy barriers vary between 0.35 and 0.8 eV, the resulting TER 
ratio remains rather insensitive and takes small values (see Figures \ref{fig10}(a) and 
(b)). Because of the practical limitation posed to the lower values of 
j$_{ON}$, these data show that it may be impractical to increase the band 
offsets above 0.8 eV.

\subsection*{Effect of dielectric-ferroelectric barrier, $U_{c}$}

\begin{figure}[htp]
	\begin{center}
		\subfigure {\label{fig11a} \includegraphics [width=2.5in,height=2.5in]{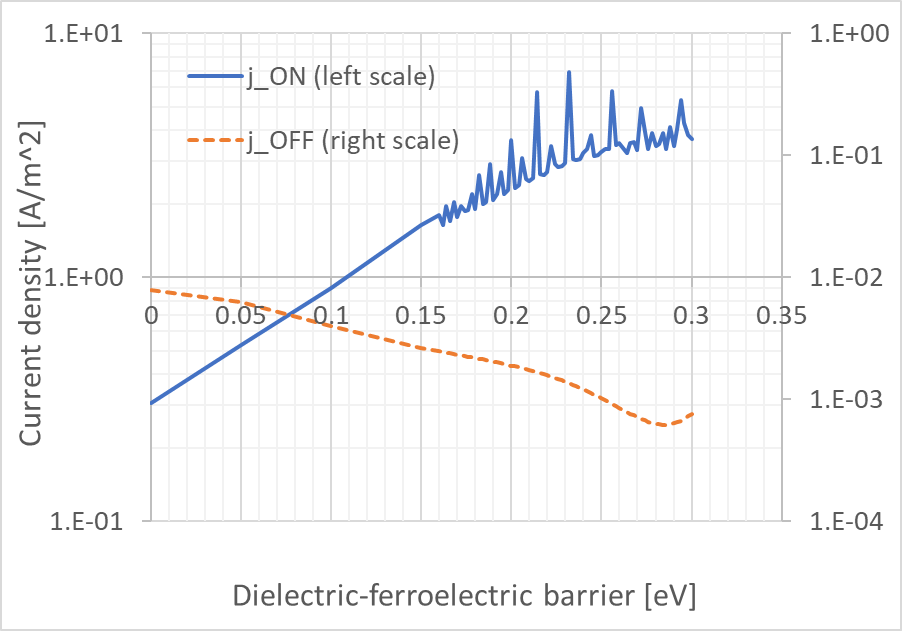}}  
		\subfigure {\label{fig11b} \includegraphics [width=2.50in,height=2.5in]{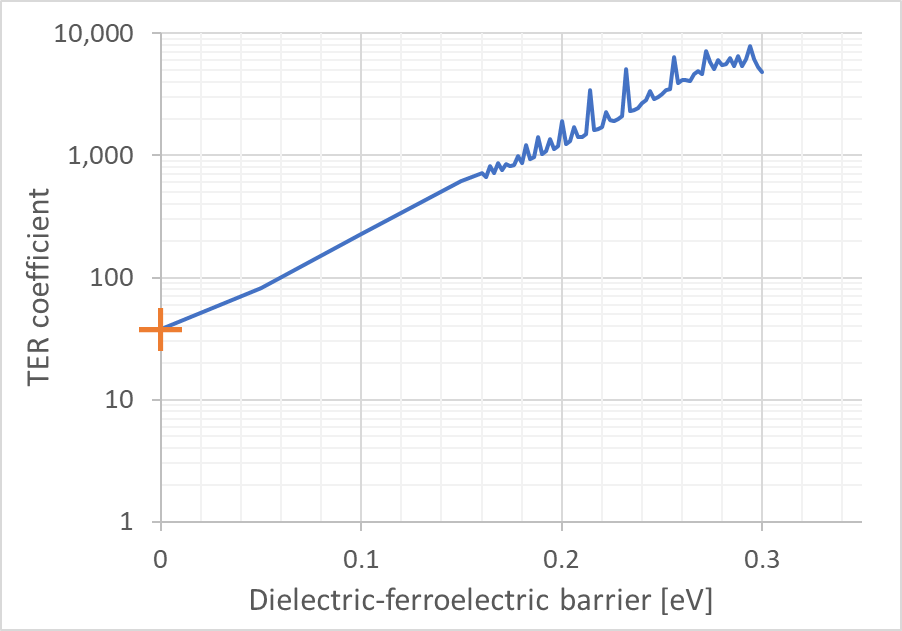}} 
	\end{center}
	\caption{(a) Current densities versus $U_{c}$ at $T$ = 300 K (b)  TER coefficient versus $U_{c}$ at $T$ = 300 K.}
	\label{fig11}
\end{figure}

\begin{figure}[H]
	\begin{center}
		\subfigure {\label{fig12a} \includegraphics [width=2.5in,height=2.5in]{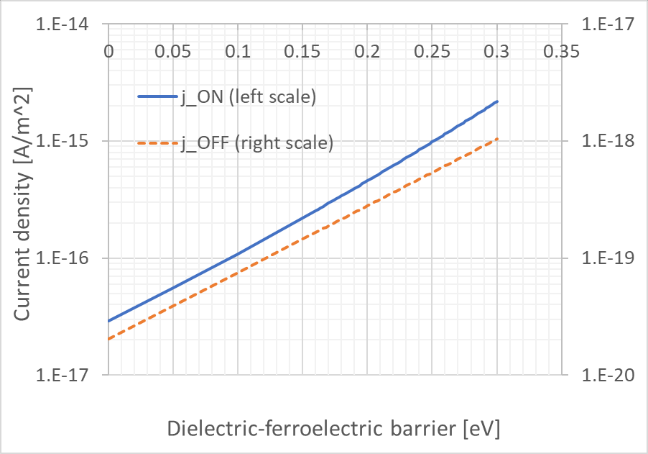}}  
		\subfigure {\label{fig12b} \includegraphics [width=2.50in,height=2.5in]{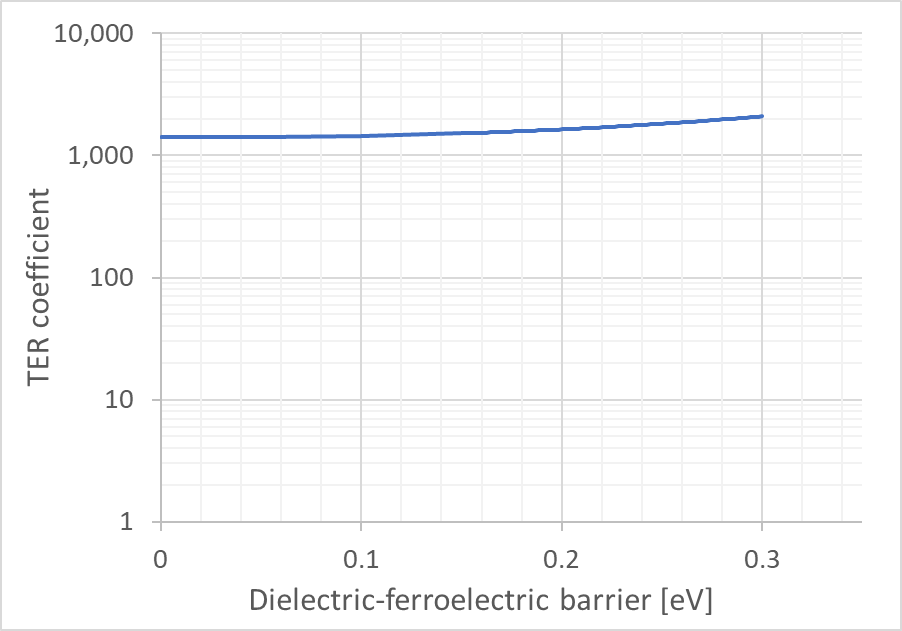}} 
	\end{center}
	\caption{(a) Current densities versus $U_{c}$ at $T$ = 50 K. The 
		oscillating behavior occurring at 300 K disappears at low T (b)  TER coefficient versus $U_{c}$ at $T$ = 50 K.}
	\label{fig12}
\end{figure}

The plots of j$_{ON(OFF)}$ versus U$_{c}$ in Figure \ref{fig11}(a) show that 
increasing the dielectric-ferroelectric barrier determines a significant 
increase of j$_{ON}$ and a moderate decrease of j$_{OFF}$, both variations 
leading to a$^{ }$three orders of magnitude increase in TER (see Figure \ref{fig11} 
(b)). Also, one may note the oscillations appearing on the upper part of 
j$_{ON}$ and TER curves. Recently, such features have been related to 
emerging resonant states at energies above the triangular barrier energy, 
where U$_{c}$ is situated, at temperatures up to 300 K \cite{Sandu2022}. It has also 
been suggested that such resonances may enhance the thermionic contribution 
to j$_{ON}$ and hence determine the increase of the TER ratio. Recalling 
that the present simulations have been performed considering ambient 
temperature, which is of interest for most applications, we have calculated 
all quantities at 50 K, and the results are plotted in Figures \ref{fig12}(a) and (b). 
Both j$_{ON}$ and j$_{OFF}$ increase by about two orders of magnitude when 
U$_{c }$ increases, whereas TER varies only between $\sim $ 1.5$ \times 
$10$^{3}$ -- 2$ \times $10$^{3}$. However, the most important result of the 
simulations at 50 K is the disappearance of the features in j$_{ON}$ an TER 
curves, which supports their assignment to the presence of resonant states 
at higher temperature, in fair agreement with the findings in \cite{Sandu2022}.

\subsection*{Effect of dielectric thickness, $t_{DE}$}
\begin{figure}[htp]
	\begin{center}
		\subfigure {\label{fig13a} \includegraphics [width=2.5in,height=2.5in]{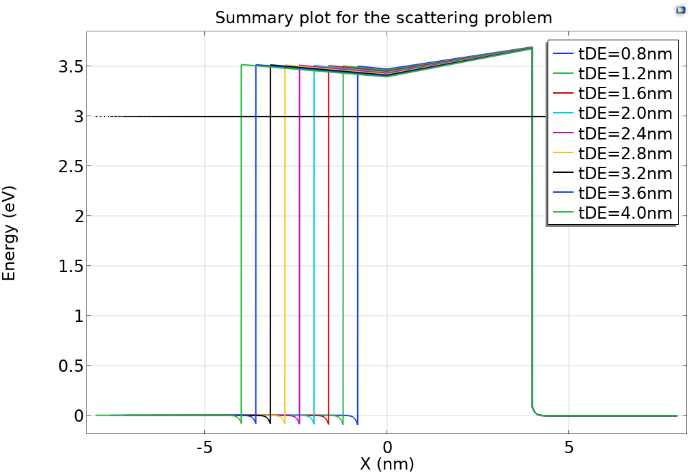}}  
		\subfigure {\label{fig13b} \includegraphics [width=2.50in,height=2.5in]{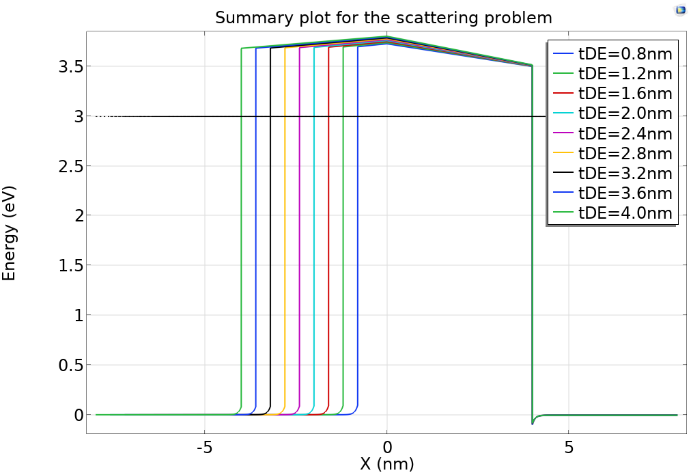}} 
	\end{center}
	\caption{(a) FTJ's barrier as function of dielectric thickness 
		($t_{DE})$, for -$P$ polarization. The average heights decrease when t$_{DE}$ increases (b)  FTJ's barrier as function 
of dielectric thickness 	($t_{DE})$, for +$P$ polarization. The average heights increase when t$_{DE}$ increases.}
	\label{fig13}
\end{figure}

In this Subsection we present the results which additionally include the 
potential barriers as function of dimensional parameter $t_{DE}$ (Figures \ref{fig13}(a) 
and (b)), when the other parameters are fixed to those of the 
SRO/STO/BTO/SRO reference system. Note that the barrier is independent of 
$T$ and only the transport through the barrier is temperature dependent.

\begin{figure}[H]
	\begin{center}
		\subfigure {\label{fig14a} \includegraphics [width=2.5in,height=2.5in]{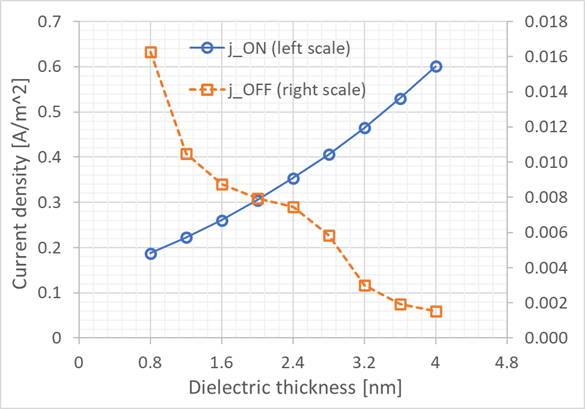}}  
		\subfigure {\label{fig14b} \includegraphics [width=2.50in,height=2.5in]{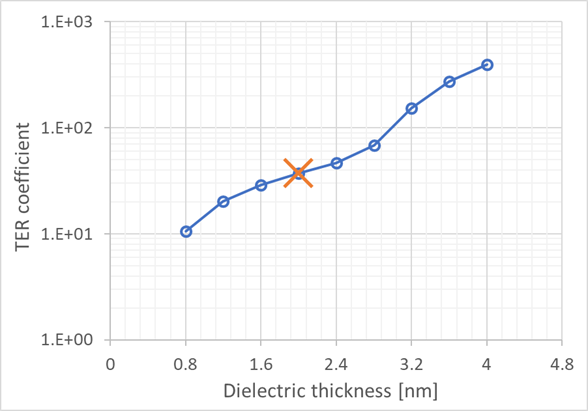}} 
	\end{center}
	\caption{(a)  Current densities as function $t_{DE}$ at $T$ = 300 K (FEM 
		simulations) (b)  TER as function of $t_{DE}$ at $T$ = 300 K (FEM 
		simulations)}
	\label{fig14}
\end{figure}

\begin{figure}[H]
\centering
\includegraphics[width=4.00in,height=3.00in]{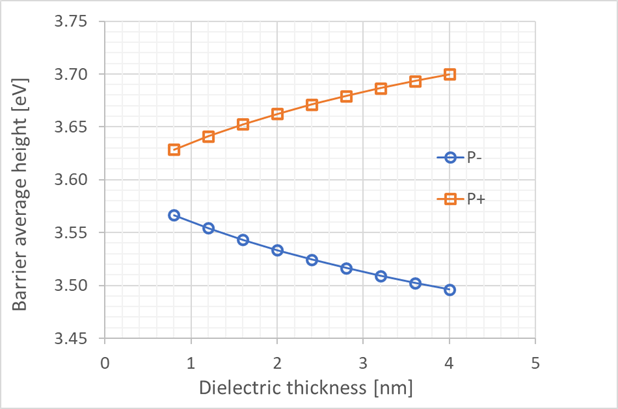}
\caption{Average barrier heights as a function of t$_{DE}$; derived 
	from barrier profiles simulated by FEM}
\label{fig15}
\end{figure}

By using the FEM-based approach presented in this paper, the current 
densities and TER coefficient were simulated at $T$ = 300 K (see Figures \ref{fig14}(a) 
and (b)). One may note an intriguing behavior related to the current density 
of the ON state, which increases when barrier width substantially increases. 
However, at the same time, the average barrier height only slightly 
decreases (Figure \ref{fig15}), which is in apparent contradiction with the direct 
tunneling theory for a rectangular barrier \cite{Simmons1963}. However, we underscore that 
within our model the temperature dependent conduction mechanisms are 
included, so this behavior should be the result of thermionic conduction 
mechanism, prevailing at 300 K. Also, these results agree with experimental 
observations \cite{Garcia2014}.

\begin{figure}[H]
	\begin{center}
		\subfigure {\label{fig16a} \includegraphics [width=2.5in,height=2.5in]{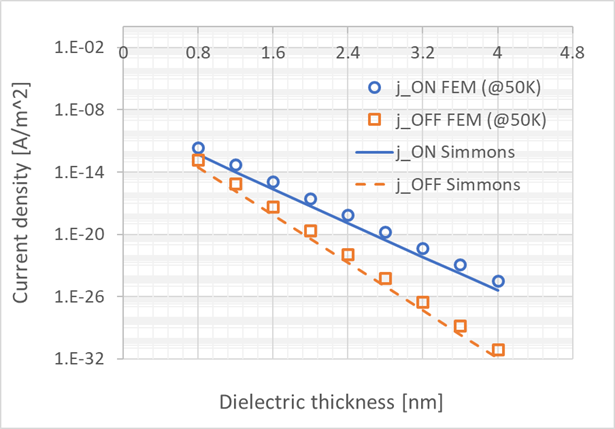}}  
		\subfigure {\label{fig16b} \includegraphics [width=2.50in,height=2.5in]{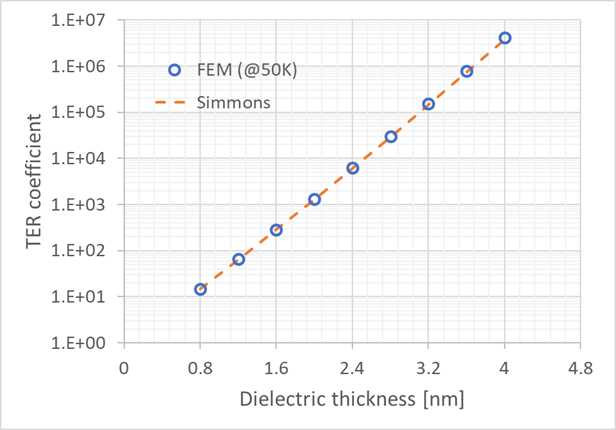}} 
	\end{center}
	\caption{(a)  Current densities as a function of t$_{DE}$ computed by FEM simulations and by Simmons equation at $T$ 
= 50 K (b) TER as a function of t$_{DE}$ computed by FEM simulations and by Simmons equation at $T $= 50 K}
	\label{fig16}
\end{figure}

To verify the statement, the barriers from Figure \ref{fig13} were averaged, and their 
heights were used in Simmons equation \cite{Simmons1963,Chapline2007} for calculating the current 
densities and TER coefficient. Because the Simmons equation is derived at 
$T$ = 0 K, we performed a FEM simulation at $T$ = 50 K (see Figure \ref{fig16}(a) and (b)), 
to allow a comparison of the results.

Thus, by comparing the results of FEM simulations and analytical 
computations (Simmons equation) one may see that the data fit quite well, 
which indicate that averaging the barriers and using Simmons equation is 
still a valid approach at T = 50 K. However, at ambient temperatures ($T$ = 300 
K), only the FEM simulations are able to predict the FTJ behavior in 
agreement with experiments, and one can note huge differences in current 
densities and TER values between 50 K and 300 K.

\section*{Discussion}

In order to substantiate the results of the various simulations performed in 
this study we mainly examine their effect on the TER ratio. First, one may 
note that the variation of the potential barriers at the 
dielectric/electrode and ferroelectric/electrode interfaces, $U_{d} $ and 
$U_{f}$, respectively, has a smaller effect, less than 15{\%} on the TER 
values, but of about seven orders of magnitude on the current densities. 
This is a non-intuitive outcome, as one could expect a major effect of the 
height and shape of $U_{d}$ and $U_{f}$ on TER, as predicted in the WKB 
approach previously \cite{Velev2016}.

Second, the group of parameters that includes the Thomas-Fermi length and 
electrode permittivity shows a rather significant effect on TER, (Figures \ref{fig7}(a) 
an (b)). However, in contrast to the two-interface FTJ systems, were the 
asymmetry due to different Thomas-Fermi lengths dictates the magnitude of 
TER \cite{Chang2017, Banerjee2019, Pantel2010}, in the composite barrier system under study, \textit{$\lambda $ / $\varepsilon $}$_{EL}$ 
variations have only a moderate impact.

Third, the parameters with really huge impact on the TER value (orders of 
magnitude) pertain to the remaining group, as follows. Thus, increasing the 
polarization from 0.1 to 0.3 C/m$^{2}$ increases TER by two orders of 
magnitude. The TER versus $P$ curve in Figure \ref{fig6}(b) follows an exponential trend, 
with a good accuracy. The electric permittivities of the dielectric and 
ferroelectric have similar effects on TER. By decreasing their values one 
can reach remarkable high TER values, especially in the case of 
\textit{$\varepsilon $}$_{DE}$ (Figure \ref{fig9}(b)). The influence of the dielectric and ferroelectric 
permittivity on the TER can be physically ascribed to the screening effect 
on the barrier potential. Consequently, the higher TER values are obtained 
at low values of dielectric and ferroelectric permittivities (Figures \ref{fig8} and \ref{fig9}).

The dielectric-ferroelectric barrier, $U_{c}$, determines a quite different 
aspect of the TER curve at energies higher than 0.15 eV, which is the 
threshold value for which resonance may accommodate in the triangular well 
at the top of the barrier. The observed oscillations are indicative of the 
fact that resonances play a significant role in the transport process at 
high temperatures. However, the oscillations disappear at lower temperature 
of the system, such as $T$ = 50 K (Figures \ref{fig13}(a) and (b)), in support of a 
resonant mechanism of electron transport at high temperature \cite{Sandu2022}. We also 
note that increasing the dielectric thickness determines large increases in 
j$_{ON}$ and TER values at 300 K, (Figures \ref{fig14}(a) and (b)), an effect that we 
ascribe to a large thermionic contribution to j$_{ON}$ via resonance states.

It is worth to highlight that in most of the studied cases at 300 K, acting 
such as to increase the TER value bears a cost on the ON state current 
density, by decreasing it, which is common wisdom in the field \cite{Garcia2014, Pantel2010}.

\section*{Conclusions}

FEM-based simulations using the Tsu-Esaki formula applied to a composite 
barrier FTJ, under external bias voltage and at room temperature, were 
performed considering a parametrized model starting from a reference one. 
The selected reference is SRO/STO/BTO/SRO system, for which a complete data 
set has been reported in literature. Using a continuum medium model, the 
current densities (for ON/OFF states) and TER coefficient were computed as a 
function of seven different material parameters, and a dimensional one (the 
dielectric layer thickness) each of them being varied in a reasonable range 
of values.

In our approach, the transport through the barrier problem was solved in 
principle exactly by using a FEM software tool, revealing that the 
temperature is important in determining the FTJ's figure of merit. The 
importance of temperature is clearly visible when comparing the results 
obtained by varying the thickness of the dielectric, at 50 K and 300 K, 
respectively, where one can notice huge differences in current densities and 
TER values. Also, it is found that analytical computations using Simmons' 
equation and averaged barrier heights predict current densities and TER 
dependences versus t$_{DE}$ in good agreement with the FEM results at 50 K. 

The results for the present case show that several configurations with large 
TER values may be predicted, but at the expense of very low current density 
in the ON state. Therefore, the FEM simulations can be more valuable than 
approximate models (direct tunneling, Fowler-Nordheim, and thermionic 
tunneling) when we are interested in the prediction of FTJ characteristics 
at temperatures close to 300 K.

We suggest that the present approach may provide useful guidelines in the 
design of preoptimized FTJs, by choosing appropriate combinations of 
materials. Also, the method may be implemented in computational schemes for 
generating training data sets for Machine Learning, if connection to 
comprehensive materials data bases is ensured.

\bibliography{sample1}

\begin{thebibliography}{10}
\urlstyle{rm}
\expandafter\ifx\csname url\endcsname\relax
  \def\url#1{\texttt{#1}}\fi
\expandafter\ifx\csname urlprefix\endcsname\relax\def\urlprefix{URL }\fi
\expandafter\ifx\csname doiprefix\endcsname\relax\def\doiprefix{DOI: }\fi
\providecommand{\bibinfo}[2]{#2}
\providecommand{\eprint}[2][]{\url{#2}}

\bibitem{Esaki1971}
\bibinfo{author}{Esaki, L.}, \bibinfo{author}{Laibowitz, R.~B.} \&
  \bibinfo{author}{Stiles, P.~J.}
\newblock \bibinfo{journal}{\bibinfo{title}{Polar switch}}.
\newblock {\emph{\JournalTitle{IBM Tech. Discl. Bull.}}}
  \textbf{\bibinfo{volume}{13}}, \bibinfo{pages}{2161} (\bibinfo{year}{1971}).

\bibitem{Zhuravlev2009}
\bibinfo{author}{Zhuravlev, M.~Y.}, \bibinfo{author}{Wang, Y.},
  \bibinfo{author}{Maekawa, S.} \& \bibinfo{author}{Tsymbal, E.~Y.}
\newblock \bibinfo{journal}{\bibinfo{title}{Tuneling electroresistance in
  ferroelectric tunnel junctions with composite barriers}}.
\newblock {\emph{\JournalTitle{Appl. Phys. Lett.}}}
  \textbf{\bibinfo{volume}{95}}, \bibinfo{pages}{052902}
  (\bibinfo{year}{2009}).

\bibitem{Garcia2014}
\bibinfo{author}{Garcia, A.} \& \bibinfo{author}{Bibes, M.}
\newblock \bibinfo{journal}{\bibinfo{title}{Ferroelectric tunnel junctions for
  information storage and processing}}.
\newblock {\emph{\JournalTitle{Nat. Commun.}}} \textbf{\bibinfo{volume}{5}},
  \bibinfo{pages}{4289} (\bibinfo{year}{2014}).

\bibitem{Velev2016}
\bibinfo{author}{Velev, J.~P.}, \bibinfo{author}{Burton, J.~D.},
  \bibinfo{author}{Zhuravlev, M.~Y.} \& \bibinfo{author}{Tsymbal, E.~Y.}
\newblock \bibinfo{journal}{\bibinfo{title}{Predictive modelling of
  ferroelectric tunnel junctions}}.
\newblock {\emph{\JournalTitle{npj Comput. Mater.}}}
  \textbf{\bibinfo{volume}{2}}, \bibinfo{pages}{16009} (\bibinfo{year}{2016}).

\bibitem{He2019}
\bibinfo{author}{He, J.}, \bibinfo{author}{Ma, Z.}, \bibinfo{author}{Geng, W.}
  \& \bibinfo{author}{Chou, X.}
\newblock \bibinfo{journal}{\bibinfo{title}{Ferroelectric tunneling through a
  composite barrier under bias voltages}}.
\newblock {\emph{\JournalTitle{Mater. Res. Express}}}
  \textbf{\bibinfo{volume}{6}}, \bibinfo{pages}{116305} (\bibinfo{year}{2019}).

\bibitem{Yang2019}
\bibinfo{author}{Yang, Q.} \emph{et~al.}
\newblock \bibinfo{journal}{\bibinfo{title}{Ferroelectric tunnel junctions
  enhanced by a polar oxide barrier layer}}.
\newblock {\emph{\JournalTitle{Nano Lett.}}} \textbf{\bibinfo{volume}{19}},
  \bibinfo{pages}{7385} (\bibinfo{year}{2019}).

\bibitem{Hwang2021}
\bibinfo{author}{Hwang, J.}, \bibinfo{author}{Goh, Y.} \&
  \bibinfo{author}{Jeon, S.}
\newblock \bibinfo{journal}{\bibinfo{title}{Effect of insertion of dielectric
  layer on the performance of hafnia ferroelectric devices}}.
\newblock {\emph{\JournalTitle{IEEE Transactions on Electron Devices}}}
  \textbf{\bibinfo{volume}{68}}, \bibinfo{pages}{22} (\bibinfo{year}{2021}).

\bibitem{Mehta1973}
\bibinfo{author}{Mehta, R.~R.}, \bibinfo{author}{Silverman, B.~D.} \&
  \bibinfo{author}{Jacobs, J.~T.}
\newblock \bibinfo{journal}{\bibinfo{title}{Depolarization fields in thin
  ferroelectric films}}.
\newblock {\emph{\JournalTitle{J. Appl. Phys.}}} \textbf{\bibinfo{volume}{44}},
  \bibinfo{pages}{3379} (\bibinfo{year}{1973}).

\bibitem{Franchini2020}
\bibinfo{author}{Franchini, G.} \emph{et~al.}
\newblock \bibinfo{journal}{\bibinfo{title}{Characterization and modeling of
  current transport in metal/ferroelectric/semiconductor tunnel junctions}}.
\newblock {\emph{\JournalTitle{IEEE Transactions on Electron Devices}}}
  \textbf{\bibinfo{volume}{67}}, \bibinfo{pages}{33} (\bibinfo{year}{2020}).

\bibitem{comsol}
\bibinfo{title}{Comsol multyphysics}.
\newblock \bibinfo{howpublished}{[Online] \url{http://www.comsol.com}}.

\bibitem{Junquera2008}
\bibinfo{author}{Junquera, J.} \& \bibinfo{author}{Ghosez, P.}
\newblock \bibinfo{journal}{\bibinfo{title}{First-principles study of
  ferroelectric oxide epitaxial thin films and superlattices: Role of the
  mechanical and electrical boundary conditions}}.
\newblock {\emph{\JournalTitle{J. Comput. Theor. Nanosci.}}}
  \textbf{\bibinfo{volume}{5}}, \bibinfo{pages}{2071} (\bibinfo{year}{2008}).

\bibitem{Stengel2011}
\bibinfo{author}{Stengel, M.}, \bibinfo{author}{Aguado-Puente, P.},
  \bibinfo{author}{Spaldin, N.~A.} \& \bibinfo{author}{Junquera, J.}
\newblock \bibinfo{journal}{\bibinfo{title}{Band alignment at
  metal/ferroelectric interfaces: Insights and artifacts from first
  principles}}.
\newblock {\emph{\JournalTitle{Phys. Rev. B}}} \textbf{\bibinfo{volume}{83}},
  \bibinfo{pages}{235112} (\bibinfo{year}{2011}).

\bibitem{Chang2017}
\bibinfo{author}{Chang, S.-C.}, \bibinfo{author}{Naeemi, A.},
  \bibinfo{author}{Nikonov, D.~E.} \& \bibinfo{author}{Gruveman, A.}
\newblock \bibinfo{journal}{\bibinfo{title}{Theoretical approach to
  electroresistance in ferroelectric tunnel junctions}}.
\newblock {\emph{\JournalTitle{Phys. Rev. Applied}}}
  \textbf{\bibinfo{volume}{7}}, \bibinfo{pages}{024005} (\bibinfo{year}{2017}).

\bibitem{Su2022}
\bibinfo{author}{Su, J.}, \bibinfo{author}{Li, J.}, \bibinfo{author}{Zheng,
  X.}, \bibinfo{author}{Xie, S.} \& \bibinfo{author}{Liu, X.}
\newblock \bibinfo{journal}{\bibinfo{title}{Integration of resonant band with
  asymmetry in ferroelectric tunnel junctions}}.
\newblock {\emph{\JournalTitle{npj Comput. Mater.}}}
  \textbf{\bibinfo{volume}{8}}, \bibinfo{pages}{54} (\bibinfo{year}{2022}).

\bibitem{Ma2019}
\bibinfo{author}{Ma, Z.~J.} \emph{et~al.}
\newblock \bibinfo{journal}{\bibinfo{title}{Enhanced tunneling
  electroresis-tance effect in composite ferroelectric tunnel junctions with
  asymmetric electrodes}}.
\newblock {\emph{\JournalTitle{MRS Communications}}}
  \textbf{\bibinfo{volume}{9}}, \bibinfo{pages}{258 -- 263}
  (\bibinfo{year}{2019}).

\bibitem{Gruverman2009}
\bibinfo{author}{Gruverman, A.} \emph{et~al.}
\newblock \bibinfo{journal}{\bibinfo{title}{Tunneling electroresistance effect
  in ferroelectric tunnel junctions at nanoscale}}.
\newblock {\emph{\JournalTitle{Nano Lett.}}} \textbf{\bibinfo{volume}{9}},
  \bibinfo{pages}{3539 -- 3543} (\bibinfo{year}{2009}).

\bibitem{Sandu2022}
\bibinfo{author}{Sandu, T.}, \bibinfo{author}{Tibeica, C.},
  \bibinfo{author}{Plugaru, R.}, \bibinfo{author}{Nedelcu, O.} \&
  \bibinfo{author}{Plugaru, N.}
\newblock \bibinfo{journal}{\bibinfo{title}{Insights into electron transport in
  a ferroelectric tunnel junction}}.
\newblock {\emph{\JournalTitle{Nanomaterials}}} \textbf{\bibinfo{volume}{12}},
  \bibinfo{pages}{1682} (\bibinfo{year}{2022}).

\bibitem{Ando1987}
\bibinfo{author}{Ando, Y.} \& \bibinfo{author}{Itoh, T.}
\newblock \bibinfo{journal}{\bibinfo{title}{Calculation of transmission
  tunneling current across arbitrary potential barriers}}.
\newblock {\emph{\JournalTitle{J. Appl. Phys.}}} \textbf{\bibinfo{volume}{61}},
  \bibinfo{pages}{1497--1502} (\bibinfo{year}{1987}).

\bibitem{Tsu1973}
\bibinfo{author}{Tsu, R.} \& \bibinfo{author}{Esaki, L.}
\newblock \bibinfo{journal}{\bibinfo{title}{Tunneling in a finite
  superlattice}}.
\newblock {\emph{\JournalTitle{Appl. Phys. Lett.}}}
  \textbf{\bibinfo{volume}{22}}, \bibinfo{pages}{562--564}
  (\bibinfo{year}{1973}).

\bibitem{Tuomisto2017}
\bibinfo{author}{Tuomisto, N.}, \bibinfo{author}{van Dijken, S.} \&
  \bibinfo{author}{Puska, M.}
\newblock \bibinfo{journal}{\bibinfo{title}{Tsu-esaki modeling of tunneling
  currents in ferroelectric tunnel junctions}}.
\newblock {\emph{\JournalTitle{J. of Appl. Phys.}}}
  \textbf{\bibinfo{volume}{122}}, \bibinfo{pages}{234301--10}
  (\bibinfo{year}{2017}).

\bibitem{Duan2020}
\bibinfo{author}{Duan, H.}, \bibinfo{author}{Fang, W.}, \bibinfo{author}{Liu,
  L.} \& \bibinfo{author}{Chen, W.}
\newblock \bibinfo{journal}{\bibinfo{title}{Theoretical study of bilayer
  composite barrier based ferroelectric tunnel junction memory}}.
\newblock {\emph{\JournalTitle{IEEE MTT-S International Conference on Numerical
  Electromagnetic and Multiphysics Modeling and Optimization (NEMO)}}}
  \bibinfo{pages}{1--3} (\bibinfo{year}{2020}).

\bibitem{Guo2020}
\bibinfo{author}{Guo, R.}, \bibinfo{author}{Lin, W.}, \bibinfo{author}{Yan,
  X.}, \bibinfo{author}{Venkatesan, T.} \& \bibinfo{author}{Chen, J.}
\newblock \bibinfo{journal}{\bibinfo{title}{Ferroic tunnel junctions and their
  application in neuromorphic networks}}.
\newblock {\emph{\JournalTitle{Appl. Phys. Rev.}}}
  \textbf{\bibinfo{volume}{7}}, \bibinfo{pages}{011304--20}
  (\bibinfo{year}{2020}).

\bibitem{Banerjee2019}
\bibinfo{author}{Banerjee, S.} \& \bibinfo{author}{Zhang, P.}
\newblock \bibinfo{journal}{\bibinfo{title}{generalized self-consistent model
  for quantum tunneling current in dissimilar metal-insulator-metal junction}}.
\newblock {\emph{\JournalTitle{AIP Advances}}} \textbf{\bibinfo{volume}{9}},
  \bibinfo{pages}{085302--12} (\bibinfo{year}{2019}).

\bibitem{Klyukin2018}
\bibinfo{author}{Klyukin, K.}, \bibinfo{author}{Tao, L.~L.},
  \bibinfo{author}{Tsymbal, E.~Y.} \& \bibinfo{author}{Alexandrov, V.}
\newblock \bibinfo{journal}{\bibinfo{title}{Defect-assisted tunneling
  electroresistance in ferroelectric tunnel junctions}}.
\newblock {\emph{\JournalTitle{Phys. Rev. Lett.}}}
  \textbf{\bibinfo{volume}{121}}, \bibinfo{pages}{056601--4}
  (\bibinfo{year}{2018}).

\bibitem{Simmons1963}
\bibinfo{author}{Simmons, J.~G.}
\newblock \bibinfo{journal}{\bibinfo{title}{Generalized formula for the
  electric tunnel effect between similar electrodes separated by a thin
  insulating film}}.
\newblock {\emph{\JournalTitle{J. of Appl. Phys.}}}
  \textbf{\bibinfo{volume}{34}}, \bibinfo{pages}{1793--1803}
  (\bibinfo{year}{1963}).

\bibitem{Chapline2007}
\bibinfo{author}{Chapline, M.~G.} \& \bibinfo{author}{Wang, S.~X.}
\newblock \bibinfo{journal}{\bibinfo{title}{Analytical formula for tunneling
  current versus voltage for multilayer barrier structures}}.
\newblock {\emph{\JournalTitle{J. of Appl. Phys.}}}
  \textbf{\bibinfo{volume}{101}}, \bibinfo{pages}{083706--10}
  (\bibinfo{year}{2007}).

\bibitem{Pantel2010}
\bibinfo{author}{Pantel, D.} \& \bibinfo{author}{Alexe, M.}
\newblock \bibinfo{journal}{\bibinfo{title}{Electroresistance effects in
  ferroelectric tunnel barriers}}.
\newblock {\emph{\JournalTitle{Phys. Rev. B}}} \textbf{\bibinfo{volume}{82}},
  \bibinfo{pages}{134105--10} (\bibinfo{year}{2007}).

\end{thebibliography}

\section*{Acknowledgements}

The authors acknowledge financial support from UEFISCDI Grant No. 
PN-III-P4-ID-PCE-2020-1985 and from the Core Program 14N/2019 MICRO-NANO-SIS 
PLUS of the Romanian Ministry of Research, Innovation, and Digitalization.

\section*{Author contributions statement}

N.P. proposed the research and supervised the project, C.T. designed and performed the calculations, T.S., O.N. and R.P. 
contributed to theory and data discussion. All authors contributed to the writing of the manuscript.

\section*{Additional information}

\textbf{Competing interests}  
The authors declare no competing interests.

\end{document}